\documentstyle[aps,preprint,epsf]{revtex}
\begin{document}
\input epsf

\title{Comparison of Strangeness
Production between A+A and p+p Reactions from 2 to 160 AGeV}
\author{J.C.~Dunlop, C.A.~Ogilvie}
\address{ Massachusetts Institute of Technology, Cambridge, MA 02139}

\maketitle

\begin{abstract}

The  measured K$^+/\pi^+$ ratios from heavy-ion reactions
are compared with the  K$^+/\pi^+$ ratios from p+p
reactions over the energy range 2-160 AGeV.
The K/$\pi$ enhancement in heavy-ion reactions is largest at
the lower energies, consistent with strangeness production in 
secondary scattering becoming relatively more important
than initial collisions near the kaon production threshold.
The enhancement decreases steadily from 4 to 160 AGeV, 
suggesting that the same enhancement mechanism of hadronic 
rescattering and decay of strings may be applicable over 
this full energy range. Based on existing data, the 
mid-rapidity K$^+/\pi^+$
ratio is predicted to be $0.27\pm0.05$ for the forthcoming Pb+Pb
reactions at 40 AGeV/c.

\end{abstract}

\vspace*{0.5cm}

Strangeness enhancement has been extensively discussed as a possible signature for the
quark-gluon plasma (QGP)\cite{Raf82}. A key question is
enhanced with respect to what?\cite{enhance}
Experiments with Si beams at 14.6 AGeV/c\cite{SiA} measured a 
K$^+/\pi^+$ ratio in heavy-ions that is four to five times larger than the
K$^+/\pi^+$ ratio from p+p reactions
at the same energy.
However in heavy-ion reactions secondary collisions often occur between 
resonant states, and the excitation energy of the resonances is
then available for 
particle production. When this mechanism is modelled in transport
calculations of heavy-ion reactions\cite{Sorge90,Pang92,Li95} 
the measured strangeness yield can be qualitatively
reproduced.
This effectively established a new baseline: strangeness is a potential signature
of the QGP if the measurements are above what one could reasonably produce from hadronic
rescattering. 

Strangeness enhancement has been also 
characterized within the context of thermal models\cite{PBM,Cley}.
One can predict the value of particle ratios, such as K/$\pi$, 
produced by a statistical system at temperature T 
and baryon chemical potential $\mu$.
The yields of particles from p+p reactions can be well 
described by such a statistical model\cite{Becattini}, 
but fitting the strangeness yields requires an
extra strangeness suppression factor, $\gamma_s$. 
The factor $\gamma_s$ scales the thermal yield of a strange hadron,
with strangeness quantum number $s$, by
$\gamma_s^{|s|}$. 
For p+p
reactions, $\gamma_s$=0.2 across a broad range of energies\cite{Becattini},
and this small value has
been interpreted as a canonical 
suppression due to the small volume of the system.
The same statistical analysis
of the measured yields from heavy-ion reactions at 160 AGeV required
less strangeness suppression ($\gamma_s$=0.6), and the analysis of the data at 10 AGeV
is consistent with no suppression  ($\gamma_s$=0.9-1.0), i.e.
the predicted strangeness yields
are in full equilibrium with non-strange hadrons\cite{Becattini}.
Therefore in the context of statistical models, strangeness enhancement in heavy-ion reactions
has been re-interpreted as a reduction in canonical 
strangeness suppression from p+p to A+A reactions.

There are at least two explanations for 
this change in suppression.
Either it is
driven by hadronic rescattering in heavy-ion collisions that helps to populate
the strange hadrons,
or a ``pre-hadronic'' QGP-like
phase possibly formed in the reaction\cite{Heinz99}
leads to rapid strangeness production.  
It is important to explore the 
beam energy evolution of these two scenarios. 
For example if the hadronic rescattering mechanism
dominates strangeness enhancement at 10 AGeV, 
how rapidly does this reduce as the beam
energy is increased? 
In particular, does the rescattering mechanism provide sufficient
enhancement to reproduce the measured strangeness data at 160 AGeV?
Or if a new mechanism is required to explain the high-energy data, 
then how does this mechanism turn-off as the beam-energy is reduced?

These questions are being addressed within transport models of heavy-ion
reactions. In order to reproduce the strangeness data at 160 AGeV, 
the transport models have had to move beyond 
the degrees-of-freedom of hadrons and strings\cite{Ehe96}, 
to either interacting strings\cite{Sorge90}, 
or new mechanisms to break di-quarks at one end of a string\cite{Cap99}. 
We take a complementary approach
and return to 
the original definition of strangeness enhancement based on experimental data, 
namely a comparison between the K$^+/\pi^+$
ratio in heavy-ion reactions and proton-proton reactions. 
By examining the  evolution of the K/$\pi$ enhancement 
from 2 to 160 AGeV\cite{Ahle98,Back99,NA49} we can address 
how rapidly the effects of hadronic rescattering change
with beam energy. Interpolating between 10 and 160 AGeV also
provides a data-based prediction for the 
K$/\pi$ ratio at the newly available beam energy of the SPS, 40 AGeV/c. 

There have been many critiques on the K$^+$/$\pi^+$ ratio as a QGP signature.
If the K$^+$/$\pi^+$ ratio is set by the
chemical properties of the system, then it
could reach similar values for both long-lived hadronic and QGP systems.
Theoretical work\cite{Heinz99} has therefore focused on whether a hadronic
system is large enough and lives 
long enough to reach its full level of strangeness production. 
Multi-strange particles potentially
offer more sensitivity to strangeness enhancement\cite{Raf99},
however if chemical equilibrium is reached within the strangeness sector, then
the yield of kaons contains the same information 
content as the yield of any strange hadron. 
Finally the K$^+$/$\pi^+$ ratio can be criticized because it is the ratio
of two signatures: strangeness enhancement and a possible increase in pion
multiplicity due to an increase in the system's entropy\cite{Gor,Kap95}.
Despite these caveats, the K$^+/\pi^+$ ratio still
provides a useful comparison between
heavy-ion and p+p data by removing to first order the 
increase in both kaon and pion yields due to system-size. 

\begin{figure}[htb]
\epsfxsize=12cm\epsfbox[30 60 520 465]{
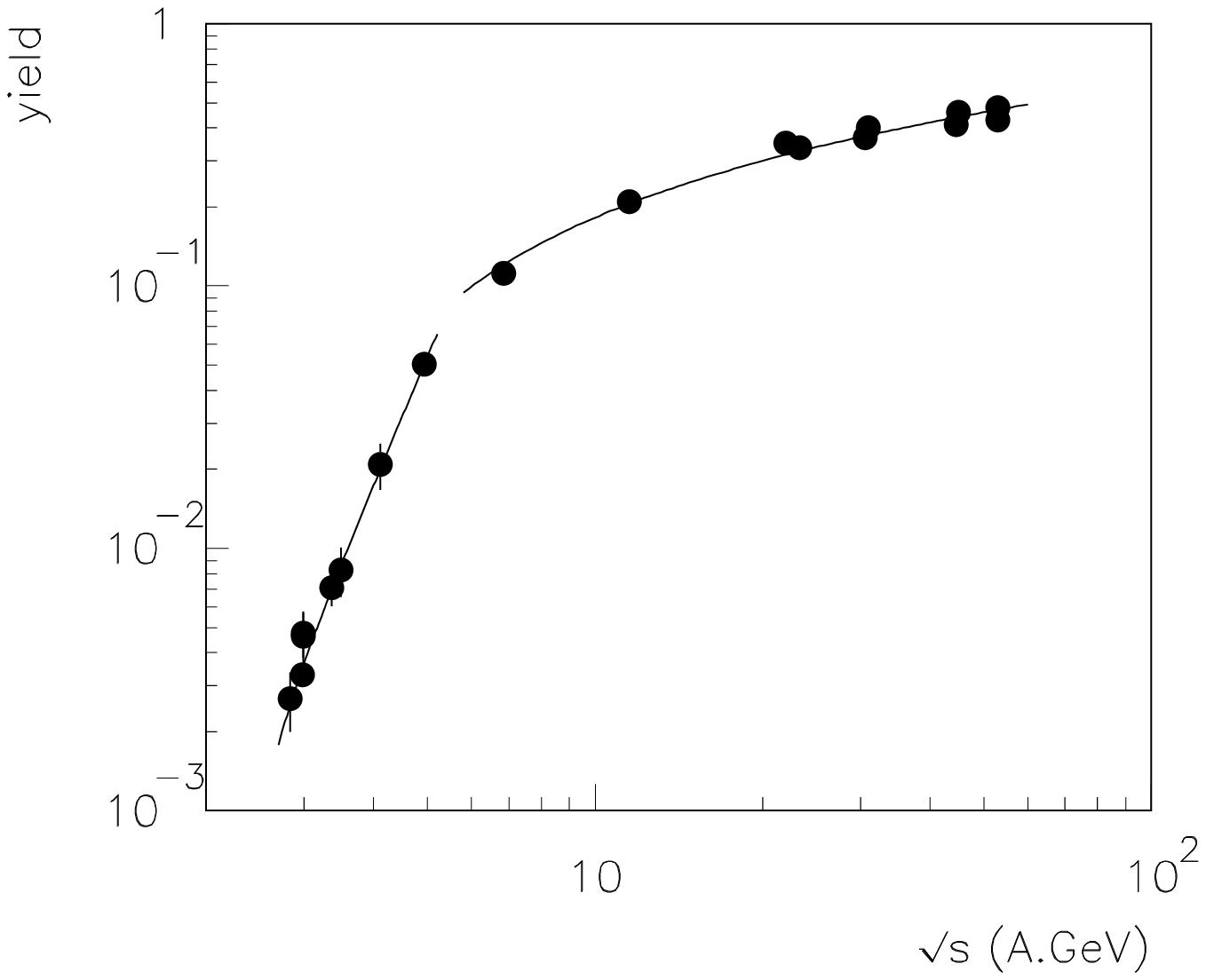}
\caption{A compilation of K$^+$ yields from p+p reactions as a function 
of s$^{1/2}$ \protect\cite{Fes79,Reed68,lb,Ant}. 
The lines are a piecewise 
parameterized fit to the data as described in the text.}
\label{fig:k_proton}
\end{figure}

Data for inclusive K$^+$ yields in p+p reactions over a broad energy
have been published 
in the literature\cite{Fes79,Reed68,lb,Ant} and are shown in
Figure~\ref{fig:k_proton}. 
No single parameterization was found that could accurately 
describe these yields over the full energy range, instead
a piecewise parameterization was used. The form of 
the low-energy part was proposed in reference \cite{Sib}
and the form of the high-energy portion has been used by e.g. Rossi 
et al.\cite{Rossi}
\begin{eqnarray}
\rm{Y_{K+}}=c_l\times(s/s_0 - 1)^{a_l} \times (s/s_0)^{b_l} \hspace*{0.2in} 
\sqrt{s_0}<\sqrt{s}<5.0 GeV
\\
\rm{Y_{K+}}=a_h+b_h\times ln(s)+c_h/\sqrt{s} \hspace*{0.2in}
6<\sqrt{s}<20 GeV
\end{eqnarray}
where s$_0=(m_p+m_K+m_\Lambda)^2$. 
The data for $\sqrt{s}<5.0$ GeV
shown in Figure ~\ref{fig:k_proton} were fit to
obtain the parameters a$_l$=0.223, b$_l$=2.196, and c$_l=0.00221$.
The data from $\sqrt{s}>6.0$ GeV shown in 
Figure~\ref{fig:k_proton} were fit to obtain the parameters
a$_h$= -0.242, b$_h$=0.089, and c$_h$=0.128.
These two parameterizations are within 10\% of each other
at $\sqrt{s}=5.5$ GeV.
It is noted that even
lower-energy p+p data from COSY exists for kaon production,
but only 6 MeV above 
the production-threshold\cite{cosy}. This 
data does not effectively constrain the parameterization 
close to $\sqrt{s}\sim$3~GeV because the measured yield of kaons from
COSY is three orders-of-magnitude 
below the yields of the
lowest energy point shown in Figure~\ref{fig:k_proton}.

\begin{figure}[htb]
\epsfxsize=12cm\epsfbox[30 60 520 465]{
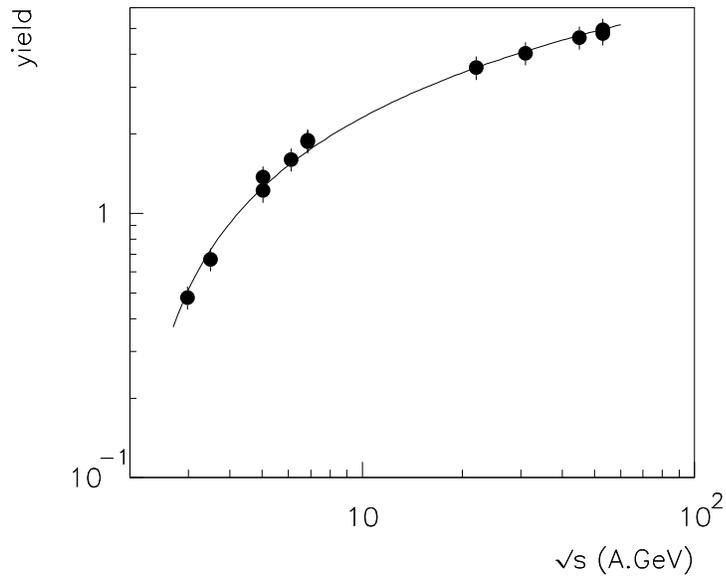}
\caption{A compilation of $\pi^+$ yields from p+p reactions as a function 
of s$^{1/2}$\protect\cite{Rossi,Ant}. 
The line is a parameterized fit that is
described in the text.}
\label{fig:pi_proton}
\end{figure}

The $\pi^+$ yields from p+p reactions shown in
Figure~\ref{fig:pi_proton} have been fit
by Rossi et al.\cite{Rossi}  with
\begin{equation}
\rm{Y_{\pi+}}=a+b\times ln(s)+c/\sqrt{s} \hspace*{0.2in}
3<\sqrt{s}<20 GeV
\label{pis}
\end{equation} We use the parameters obtained by Rossi et al.\cite{Rossi};
a= -1.55, b=0.82, and c=0.79. It is estimated that the 
systematic uncertainty of both the parameterized kaon and pion yields is 10\%,
but becomes larger towards the ends of the fitted ranges.

\begin{figure}[htb]
\epsfxsize=12cm\epsfbox[30 60 520 465]{
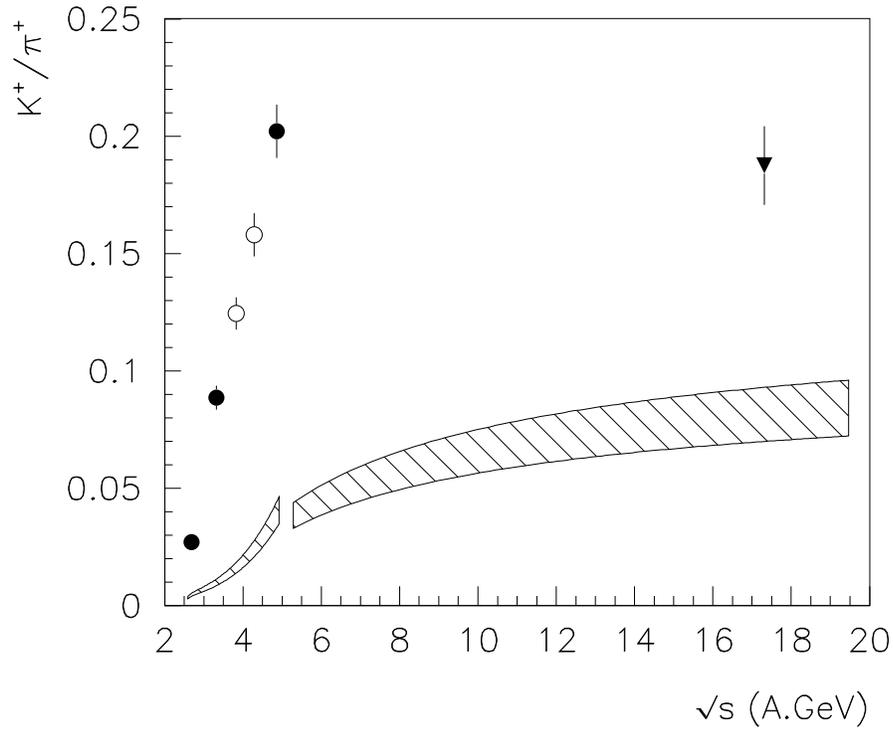}
\caption{The ratio of dN/dy for K$^+$/$\pi^+$ at mid-rapidity in
central Au+Au and Pb+Pb reactions as a function 
of the initial available energy. 
The filled circles are from E866, the open circles are from E917 and the
triangle is from NA49.
The hashed region is the K$^+$/$\pi^+$ ratio from the parameterized K and 
$\pi$ yields from p+p reactions (see text for details). The hashed region
covers $\pm1\sigma$ around the p+p K$^+$/$\pi^+$ ratio.
}
\label{fig:kpi}
\end{figure}

The K$^+$ yield from p+p reactions 
increases faster with
beam energy than the $\pi^+$ yield, such that
the K$^+/\pi^+$ ratio from p+p reactions 
(hashed region in Figure~\ref{fig:kpi}) 
increases steadily throughout the energy range. At higher energies the ratio
tends towards a value of K$^+/\pi^+$=0.08.

In heavy-ion reactions data on the 
K$^+/\pi^+$ ratio at mid-rapidity available from central Au+Au reactions
at 2, 4, 6, 8, and 10.7 AGeV\cite{Back99}, and from central Pb+Pb reactions at 158 AGeV\cite{NA49}.
These data are shown in  Figure~\ref{fig:kpi}.
The K$^+/\pi^+$ ratio increases steadily from 
0.0271$\pm0.0015\pm0.0014$ at 2~AGeV to 
0.202$\pm0.005\pm0.010$ at 10.7~AGeV\cite{Back99}.  
The measured ratio K$^+/\pi^+ = 0.19\pm 0.01$ from 
Pb+Pb collisions at 157~AGeV/c\cite{NA49} is comparable to the ratio
from Au+Au reactions at 10.7~AGeV.
This suggests that either the ratio saturates or that a maximum exists in
the K$^+$/$\pi^+$ from heavy-ion reactions at 
energies between the AGS and SPS. 
At all beam energies the K$^+/\pi^+$ ratio from heavy-ion reactions
is larger than in p+p reactions.
It is noted  that the data
from A+A are measured at 
mid-rapidity whereas the p+p results are integrated over the full
phase space. 
As an estimate of the level of the difficulties this might cause, 
in Au+Au reactions at 10.7~AGeV the 
mid-rapidity K$^+$/$\pi^+$ ratio is 0.202$\pm0.005\pm0.010$\cite{Back99} and 
is within a few percent of the value obtained by integrating over a
broader rapidity range of $0.6<y<2.0$ where 
K$^+$/$\pi^+=0.197\pm0.003\pm0.010$\cite{Ahle98}.
  
\begin{figure}[htb]
\epsfxsize=9.5cm\epsfbox[30 60 520 495]{
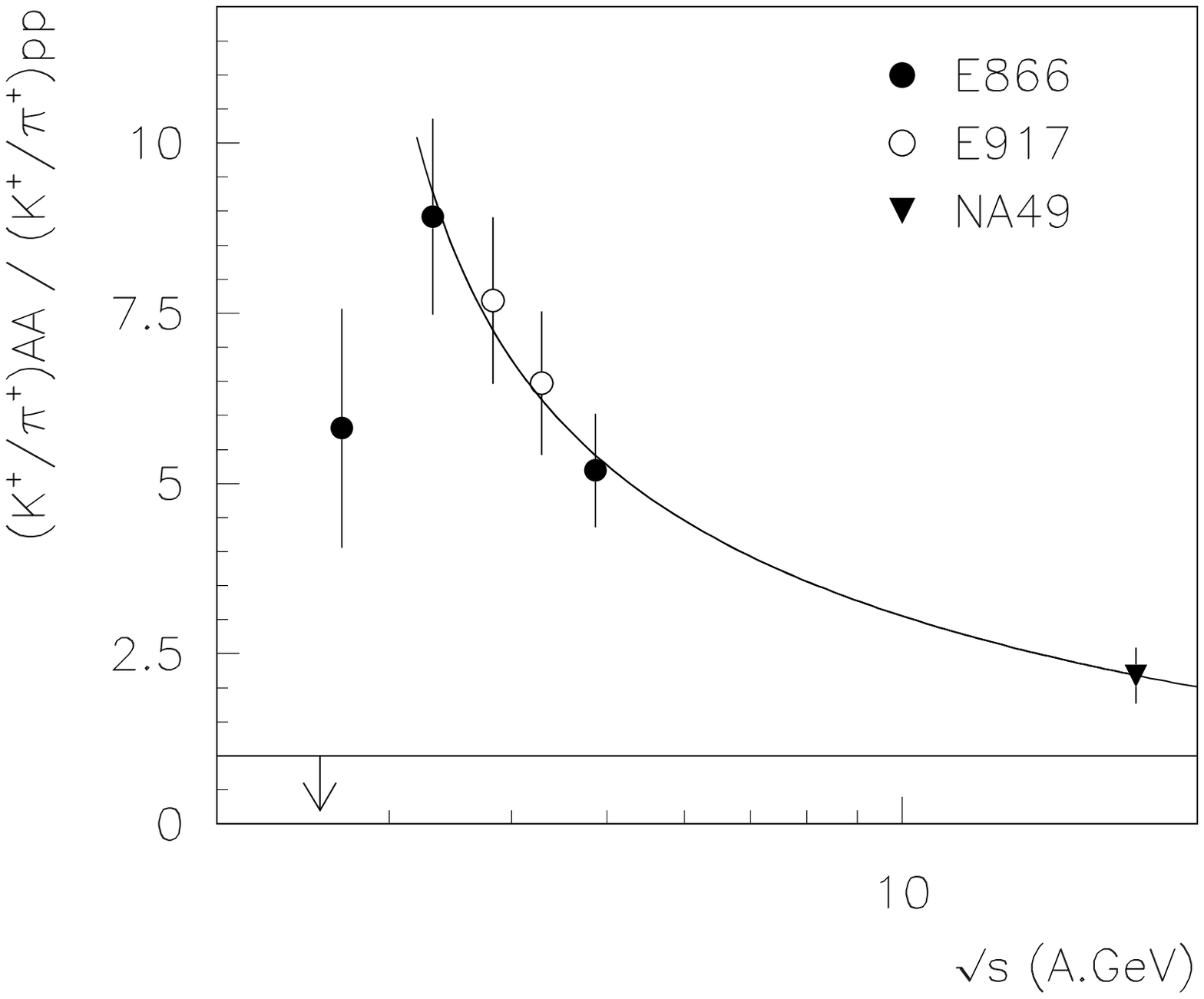}
\caption{The double ratio K$^+$/$\pi^+$ at mid-rapidity from central
Au+Au reactions divided by K$^+$/$\pi^+$ of total yields from p+p
reactions as a function 
of the initial available energy.
The errors include both statistics and a 10\%
systematic uncertainty in the parameterized kaon and pion yields
from p+p reactions. These systematic errors increase to 20\% at the lowest
beam energy (2 AGeV).
The arrow indicates the threshold energy for producing K$^+$ in a 
p+p reaction, the horizontal line is
an enhancement of one, and 
the hyperbolic line is a fit to the data (equation 4).
}
\label{fig:enhance}
\end{figure}

The measured heavy-ion K$^+/\pi^+$ ratio
divided by the p+p K$^+/\pi^+$ ratio calculated using
equations 1-3 is shown in Figure~\ref{fig:enhance}.
This double ratio is referred to in this work as the K$^+/\pi^+$ enhancement.
The enhancement is smallest at the highest beam 
energy at the SPS.
At low beam energies the K$^+/\pi^+$ enhancement in Au reactions
is likely to be caused by secondary hadron collisions.
The increase in enhancement at lower energies suggests that  
as the beam energy is reduced towards the threshold for kaon
production, secondary collisions   
increase in relative 
importance
compared to initial collisions.
At beam energies below the kaon threshold the double ratio is, by definition,
infinite.

Any increase in pion 
absorption in heavy-ion reactions  as the beam energy
decreases 
would also contribute to the enhancement of the K$^+$/$\pi^+$ ratio.
At 10.7~AGeV  
four-fifths of the K$^+$/$\pi^+$ enhancement 
is due to an increase in kaon yield, since
K$^+$ production per collision participant
in central Au+Au reactions was measured\cite{Ahle98} to be
four times larger than for nucleon-nucleon reactions.
It is not clear why the enhancement apparently decreases from 4 to
2 AGeV, though it is in this region that the parameterized
pion yields from p+p reactions are not well constrained.

The increase in importance of 
secondary collisions as the beam energy approaches a 
production threshold 
approaches is counter intuitive. Most of these secondary 
collisions occur at 
values of $\sqrt{s}$ that are lower than the $\sqrt{s}$ available in
initial nucleon-nucleon collisions and hence the 
production of kaons in a single secondary collision is 
lower than the production of kaons in a single primary collision. However in 
heavy-ion reactions the large number of secondary reactions  
compensates for this and, 
in total, produce more kaons than the initial nucleon-nucleon collisions.

The decrease in the enhancement from 4 to 160 AGeV provides a natural
way to view the existence of a maximum in the 
heavy-ion K$^+/\pi^+$ ratio. 
The K$^+$/$\pi^+$ ratio from p+p reactions increases 
in this energy range. If this is coupled with a heavy-ion reaction mechanism 
that causes the K$^+$/$\pi^+$ enhancement to fall, then the 
K$^+/\pi^+$ ratio in heavy-ion reactions must have a maximum
as a function of beam energy.

There is a large gap in the data between 10 AGeV and the 
SPS energy of 160 AGeV. However the enhancement at the SPS
is consistent with a smooth continuation of the decrease in enhancement
from the AGS energies.
To demonstrate this, the enhancement from 4-160 AGeV can be fit with
\begin{equation}
\frac{(K^+/p^+)_{AA}}{(K^+/p^+)_{pp}} = \frac{a}{(\sqrt{s}-\sqrt{s_0})^b}
\label{hype}
\end{equation}
with two free parameters a=8.2 and b=0.49
and three degrees-of-freedom ($\sqrt{s_0}$ is the threshold for 
K$^+$ production in p+p reactions).
This fit is shown as a solid line in figure~\ref{fig:enhance}.
Because both the  SPS and AGS enhancement data can be empirically
fit with the same decreasing function, it is possible
that a qualitatively similar reaction mechanism for strangeness
enhancement is present at both AGS and SPS energies.

This argument is far from establishing that the same 
enhancement mechanism $is$ at work over this full energy range.
There are many examples of strong-interaction physics where particle
production smoothly increases with beam energy, but the reaction mechanism
evolves between two scenarios. For example, the charged particle multiplicity 
steadily increases as the beam
energy is increased from a region where the data can be modeled by
the excitation and breaking of strings to higher
energies where a description of the data requires the
fragmentation of mini-jets\cite{Wang}.
However in the case of heavy-ion reactions,
there is the possibility of forming a QGP which might
be observable as
distinct changes in the characteristics of particle production
with increasing beam energy.

Whether the smooth decrease of the
K$/\pi$ enhancement continues between AGS and SPS energies
will be checked  by forthcoming measurements of Pb+Pb collisions at 
40 AGeV/c ($\sqrt{s}$=8.8 AGeV). 
From equation~\ref{hype}, the interpolated
K$^+/\pi^+$ enhancement at 40 AGeV/c is 3.3$\pm$0.5. Multiplying this
enhancement by the parameterized K$^+/\pi^+$ 
from p+p reactions (equations 1-3), 
we can make the prediction of a strikingly large
K$^+/\pi^+=0.27\pm0.05$ for Pb+Pb reactions at 40 AGeV/c. 

If the measured result is
below this value, then this would imply that 
a minimum exists in the strangeness enhancement.
A minimum would logically require 
the existence of an additional mechanism for strangeness production
at the highest SPS energy (160AGeV). 
A similar speculation can be put forward for the
forthcoming RHIC results, which will measure an excitation function
of Au+Au collisions between approximately  $50 <\sqrt{s}< 200$ AGeV.
An observation of a 
minimum in the combined AGS-SPS-RHIC excitation function of
strangeness enhancement would lead to the model-independent conclusion of
an additional source of 
strangeness production turning on at some beam energy, 
potentially driven by the quark-gluon plasma. 

In summary, the existing heavy-ion 
K$^+/\pi^+$ data have been compared with the data from p+p
reactions over the energy range 2-160 AGeV.
The K/$\pi$ enhancement is largest at
the lower energies, consistent with strangeness production in 
secondary scattering becoming relatively more important
than initial collisions near the kaon production threshold.
The enhancement decreases steadily from the AGS to the SPS. The AGS data
sets the rate of decrease for the K/$\pi$ enhancement due to the hadronic
rescattering and the formation of strings. Since the AGS and SPS
data can be both fitted with a smooth decrease in the 
strangeness enhancement, the data are consistent with 
the same enhancement mechanism at work over this full energy range.
Key to confirming or excluding this possibility is the
SPS measurement at 40 AGeV/c and the forthcoming RHIC experiments
at higher energies.
  
Discussions with M. Tannenbaum, B. Mueller and
G.S.F. Stephans are gratefully acknowledged. This work was supported by DOE.

\clearpage

\clearpage

\clearpage

\begin{thebibliography}{9}
\bibitem{Raf82}  J. Rafelski and B. M\"uller, Phys. Rev. Lett. {\bf 48}, 1066 
(1982). 
\bibitem{enhance} T. Abbott et al., Phys. Rev. Lett. {\bf 64}, 847 
(1990), O. Hansen, Comm. Nucl. Part. Phys. {\bf 20}, 1 (1991).
\bibitem{SiA} T. Abbott et al., Phys. Lett. {\bf B291} 341, (1991),
 T. Abbott et al., Phys. Rev. C {\bf 50}, 1024 (1994)
\bibitem{Sorge90} H. Sorge, H. St\"ocker and W. Greiner
, Ann. Phys. (NY) {\bf 192}, 266 (1989).
\bibitem{Pang92} Y. Pang, T.J. Schlagel, and S.H. Kahana,
Phys. Rev. Lett. {\bf 68}, 2743 (1992).
\bibitem{Li95} B-A. Li and C.M. Ko, Phys. Rev. C {\bf 52}, 2037 (1995).
\bibitem{PBM} P. Braun-Munzinger et al, Phys. Lett. {\bf B344} 43 (1995). 
\bibitem{Cley} J. Cleymans et al., Z. Phys. C {\bf 74}, 319 (1997)
\bibitem{Becattini} F.Becattini, J. Phys. G {\bf 25}, 287 (1999),
 F. Becattini et al., Nucl. Phys. {\bf A638}, 403 (1998).
\bibitem{Heinz99} U. Heinz, QM99 nucl-th 9907060, to be published Nucl. Phys. A
\bibitem{Ehe96} W. Ehehalt and W. Cassing, Nucl. Phys. {\bf A602}, 449
  (1996).
\bibitem{Cap99} A. Cappella et al.,  Phys. Lett. {\bf B459} 27, (1999), 
\bibitem{Ahle98} L. Ahle et al., E866 Collaboration, Phys. Rev. C {\bf 58}, 3523 (1998).
\bibitem{Back99} B. Back et al.,  E866, E917 Collaborations, to be published, nucl-ex/9910008.
\bibitem{NA49} F. Sikler, NA49 Collaboration, Quark Matter 1999, to be published Nucl. Phys. A
\bibitem{Fes79} H. Fesefeldt et al., Nucl. Phys. {\bf B147} (1979) 317.
\bibitem{Reed68} J.T. Reed et al, Phys. Rev. {\bf 168}, 1495 (1968).
\bibitem{lb} Numerical Data and Functional Relationships
in Science and Technology, Landolt-B\"ornstein New Series I/12B,
Berlin Springer-Verlag (1993)
\bibitem{Ant} M. Antinucci et al., Nuovo Cimento Lett. {\bf 6}, 121 (1973).
\bibitem{cosy} J.T. Balewski et al., Phys. Lett. B {\bf 388}, 859 (1996)
and D.Grzonka Nucl. Phys. {\bf A631}, 262c (1998). 
\bibitem{Sib} A.A. Sibirtsev, Nucl. Phys. {\bf A604}, 455 (1996).
\bibitem{Rossi} A.M. Rossi et al., Nucl. Phys. {\bf B84}, 269 (1975).  
\bibitem{Raf99} J. Rafelski, J. Phys. {\bf G25}, 451 (1999).
\bibitem{Gor} M.I. Gorenstein et al., Phys. Lett. {\bf B281}, 197 (1992).
\bibitem{Kap95} J.I. Kapusta, A.P. Vischer, Phys. Rev. C 
{\bf 52}, 2725 (1995).
\bibitem{Wang} X.-N. Wang, Phys. Rep. {\bf 280}, 287 (1997).

\end{thebibliography}
\end{document}